\documentclass[twocolumn]{aastex7}

\usepackage{makecell, amsmath, amsfonts, amssymb,stix}
\usepackage{cleveref}
\renewcommand\micron{\textmu{m}}
\newcommand{\Autoref}[1]{%
  \begingroup
  \def\figureautorefname{Figure}%
  \def\tableautorefname{Table}%
  \def\sectionautorefname{Section}%
  \def\subsectionautorefname{Subsection}%
  \def\equationautorefname{Equation}%
  \autoref{#1}%
  \endgroup
}

\shorttitle{Uranian Rings and Small Moons with JWST}
\shortauthors{Belyakov et al.}

\begin{document}

\title{JWST Reveals the Kuiper Belt-like Compositions of the Uranian Rings and Small Moons}

\author[orcid=0000-0003-4778-6170]{Matthew Belyakov} 
\affiliation{Division of Geological and Planetary Sciences, California Institute of Technology, Pasadena, CA 91125, USA}
\email[show]{mattbel@caltech.edu}

\author[orcid=0000-0002-7451-4704]{M. Ryleigh Davis}
\affiliation{Division of Geological and Planetary Sciences, California Institute of Technology, Pasadena, CA 91125, USA}
\email{rdavis@caltech.edu}  

\author[orcid=0000-0001-5683-0095]{Zachariah Milby}
\affiliation{Division of Geological and Planetary Sciences, California Institute of Technology, Pasadena, CA 91125, USA}
\email{zmilby@caltech.edu}  

\author[orcid=0000-0001-9665-8429]{Ian~Wong}
\affiliation{Space Telescope Science Institute, 3700 San Martin Drive, Baltimore, MD 21218, USA}
\email{iwong@stsci.edu}

\author[orcid=0000-0002-8255-0545]{Michael E. Brown}
\affiliation{Division of Geological and Planetary Sciences, California Institute of Technology, Pasadena, CA 91125, USA}
\email{mbrown@caltech.edu}

\begin{abstract}
We present the JWST NIRSpec 1--5 \micron{} spectra of the Uranian rings and four ringmoons: Puck, Perdita, Portia, and Juliet. We find that the compositions of  the rings and moons strongly resemble water-type Kuiper belt objects, showing deep 3.0 \micron{} O--H and water ice features as well as the CO$_2$ band at 4.27 \micron{}. The volatiles show a radial trend of decreasing volatile abundance at closer radial distance to the planet, consistent with recent estimates of the densities of the innermost moons. Given the observed compositional signatures, we suggest two ways to reconcile the history of the small inner moons with that of the regular ones. The first is that the Uranus-tilting collision produced a disk from an impactor formed out of material similar to that which comprises water-rich Kuiper belt objects, and the ringmoons are small undifferentiated remnants of that process. Alternatively, the rings and ringmoons postdate the formation of the regular moons, and formed from the later tidal disruption of a several hundred kilometer sized undifferentiated water-type Kuiper belt object.
\end{abstract}

\section{Introduction} 
All of the Solar System's giant planets host a set of rings and small inner moons sometimes termed ``ringmoons''. These ring-and-moon systems exhibit complex dynamical behavior, with the structure of the rings simultaneously shaped by resonant interactions with both the ringmoons and global-scale normal mode oscillations of the host planets' interiors \citep{Nicholson2018, Showalter2020}. The small inner moons are themselves also significantly affected by the dense network of mean-motion resonances they create \citep{French2015AJ, Cuk2022AJ}. The outcome of these many interactions is hypothesized to be a cycling of material: the rings and small moons exchange material when moons collide, forming rings which then re-coalesce into a new generation of moons \citep{French2015AJ,Charnoz2018prs,Hesselbrock2019AJ}. This chaotic dynamical evolution may have erased the original configuration of these systems, leaving composition as the key probe into the origin of the rings and small inner moons of the giant planets.

Uranus presents a particularly difficult challenge in ascertaining the origin of the planet's satellite system as a whole, as it is situated in the equatorial plane set by the planet's 98 degree obliquity. Aligning the satellites with a tilted planet requires either the destruction of the original satellites and re-accretion along Uranus' equatorial plane \citep{Morbidelli2012Icar}, their formation out of a disk of disrupted cometary material subsequent to the planet-tilting impact \citep{Crida2012Sci}, or spin-orbit coupled dynamics that tilted the system as a whole \citep{Saillenfest2022A&A}. Hypotheses involving formation of satellites out of an impact disk mixing Uranus and impactor material have been strongly disfavored by the elevated deuterium-to-hydrogen (D/H) ratio on the system's large regular satellites \citep{pnasbrown}. In this context, the composition of the Uranian rings and small moons may provide hints for the formation history of the larger satellites.

Uranus hosts a dense system of 9 narrow rings and 14 small moons, alongside several diffuse dusty rings including the $\mu$ and $\nu$ rings. The discovery of the first rings around Uranus came from occultations measured by the Kuiper Airborne Observatory in 1977 \citep{1977Natur.267..328E}, with additional rings identified in 1981 \citep{elliot1981orbits}. The Voyager 2 fly-by identified 11 of the known small moons \citep{Smith1986Sci, Porco1987AJ} and obtained resolved images of Puck, the largest of the inner satellites. Subsequent observations by the Hubble Space Telescope (HST) detected the outermost moon of the ring-moon system, Mab, along with Cupid. Recently, James Webb Space Telescope (JWST) Near-infrared Camera (NIRCam) imaging discovered the smallest satellite yet, S/2025 U1 \citep{ElMoutamid2025}.

\begin{figure*}
    \centering
    \includegraphics[width=\textwidth]{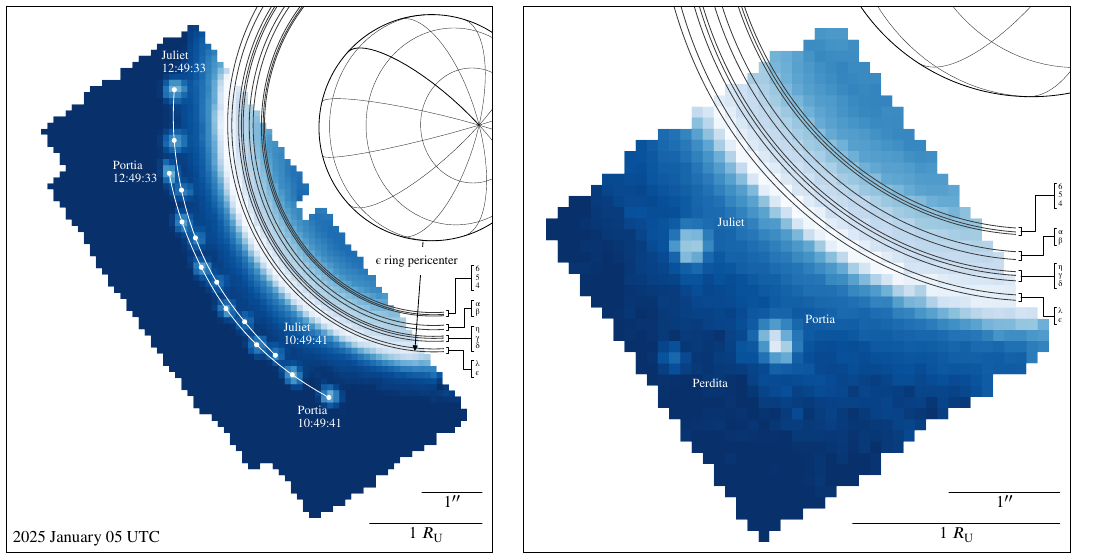}
    \caption{JWST NIRSpec detector images at 2.48 \micron. The wireframe depicts the state and orientation of Uranus and its rings at the midpoint of the exposure(s). \textit{Left:} The seven dithers targeting Portia stitched together relative to the centroid of Uranus. Labels show the times and positions of the moons during the first and last dithers. Dots show the centroids of each satellite and the connecting lines illustrate their orbital paths. \textit{Right:} The second dither of our Portia observations, with a brightness stretch chosen to reveal the faint moon Perdita as well as Portia and Juliet. The width of the PSF and relative brightness of the $\epsilon$ ring obscure the structure of the ring system as seen in the NIRCam images from \cite{Hedman2025PSJ}.}
    \label{fig:jwstobs}
\end{figure*}

Characterization of Uranus' ring-moon system has proven difficult due to the combination of scattered light from the planet and the small size of the moons. Prior to JWST, occultation studies, Voyager imaging, and HST/ground-based spectrophotometry revealed only the most basic compositional information about the rings and moons. The albedos of the ring particles and moons are measured to be $\sim$0.05 \citep{Smith1986Sci}, much lower than Saturn's icy rings or the adjacent regular Uranian satellites, and instead are more similar to other outer Solar System small bodies. HST spectrophotometry found that Puck and the other small moons have slightly red visible spectra, with weak 2.0 \micron{} absorption features attributed to water ice \citep{Karkoschka2001Icar}. By contrast, spectroscopy of the $\epsilon$ ring, which is the brightest of the set of rings, has placed strong upper limits on the 2.0 \micron{} water ice band \citep{deKleer2013Icar}.

Infrared spectrophotometry by JWST NIRCam has provided an updated view of the compositions of the Uranian small moons and rings by detecting the 3.0 \micron{} feature common to water ice and hydrated minerals on all observed moons and rings. Preliminary work by \cite{Belyakov2024PSJ} was the first to report the 3.0 \micron{} feature on the surfaces of Puck, Portia, Belinda, Cressida, and Juliet. This feature is of similar depth to those seen on water-type Kuiper belt objects (KBOs) \citep{PinillaAlonso2025NatAs, Belyakov2025PSJ, Holler2025RNAAS,Wongbluebinary, Wong2025PSJ}, indicating that water ice and/or hydrated minerals are the source of the band. Additional images and improved processing allowed \cite{Hedman2025PSJ} to extend detections of the 3.0 \micron{} band to Bianca, Desdemona, and Rosalind. The outermost moon, Mab, appears to have a very blue slope, suggesting its surface may be very water-rich, unlike the rest of the small moons. Spectrophotometry of the rings found that their 3.0 \micron{} band depths were 25\% lower than those of the moons, hinting at compositional heterogeneity between the rings and moons. A nascent trend of 3.0 \micron{} band depth increasing with radial distance has also been reported \citep{Hedman2025PSJ}.

In this Letter, we present the first JWST NIRSpec 1--5 \micron{} spectra of the moons Puck, Perdita, Portia, Juliet and the Uranian rings. \Autoref{sec:methods} discusses the observations and data reduction, describing the method used to remove the contribution of scattered light from Uranus in the spectra. \Autoref{sec:spec} shows the spectra of the small moons and rings in the context of the water-rich KBOs, finding significant similarity between the two sets of objects. Additionally, the band depths of key features as a function of radial distance are shown. Finally, \Autoref{sec:discussion} explores the implications of the spectroscopic findings for the formation and evolutionary history of the inner moons and rings.

\section{Methods} \label{sec:methods}


Near-infrared spectroscopy of the Uranian rings and ringmoons was carried out with the integral field unit (IFU) of the Near-Infrared Spectrograph (NIRSpec) as part of JWST Program 4645 (PI: M. Belyakov), which observed the small and irregular satellites of Uranus and Neptune. Two visits were made to the Uranian system: the first to obtain the spectrum of Puck, and the second to observe Portia. By observing Portia, we simultaneously acquired the spectrum of the rings along with serendipitous observations of two other moons, Perdita and Juliet, all of which were present in the IFU across all 7 dithers, as shown in \autoref{fig:jwstobs}. All observations were taken in PRISM mode ($R\sim$30 to 300), with the NRSIRS2RAPID readout pattern selected to reduce correlated detector readnoise.

Our data processing involved several changes from typical JWST NIRSpec Solar System data processing. We began with the uncalibrated \texttt{.uncal} files, which we processed in part using the \texttt{jwstspec} pipeline described in \citet{Wong2024PSJ}. We used reference files drawn from context \texttt{jwst\_1581.pmap} of the JWST Calibration Reference Data System along with version 3.0.0 of the standard JWST calibration pipeline \citep{jwst}. For the ramp fitting step, we switched the method used from the typical least squares fit to a maximum likelihood algorithm as described in \cite{Brandt2024PASP}. The primary benefit of using a likelihood-driven algorithm is that it provides additional safeguards against cosmic rays or any other jumps in the ramp sampling. For $1/f$ noise correction, we used the Stage 2 \texttt{clean\_flicker\_noise} step, with the following settings: mask science pixels, fit each channel individually, and use a median fit rather than the pipeline-suggested FFT method. Bleed onto non-science parts of the detector caused by Uranus leads the FFT fit to introduce significant artifacts throughout the cube. We made two changes to the default settings at the \texttt{cube\_build} step: First, for all objects, we set the output type to ``multi'', thus building the data cube with a non-linear wavelength solution. The use of the non-linear wavelength solution mitigates fringing effects which are apparent in all of our JWST observations at short wavelengths where Uranus-shine is strongest. Additionally, sampling with the non-linear wavelength solution provides the correct resolution for measuring and reporting band centers of spectral features. Second, when reducing the data for point sources, we used the exponential modified-Shepard method (EMSM) for weighting the transformation between the 2D-detector pixels and the 3D cube, as opposed to the default drizzle algorithm \citep{Law2023AJ}. While EMSM degrades the structure of the NIRSpec PSF compared to drizzle \citep{Beck2025jwst}, it provides superior signal-to-noise ratio for faint point sources. Finally, and again only for the small moons, we built the cube along detector coordinates. For extracting spectra of the ring, we built standard cubes with drizzle weighting and alignment to equatorial sky coordinates. 

Given the Stage 2 bias-corrected, flat-fielded, spatially-rectified, wavelength-calibrated, and flux-calibrated IFU data cubes for each dithered exposure (shown in \autoref{fig:jwstobs}), we extracted the spectra of the moons and rings. As a prerequisite, we obtained coordinates of the Uranian rings at the midpoint time of each dither in order to facilitate scattered light subtraction. This step was complicated by the non-negligible inclinations and eccentricities of many of the Uranian rings. We used the ring orbital parameters measured during the Voyager flyby \citep{French1988}. For each dither, we mapped the ring positions onto the sky in right ascension and declination using the SPICE observation geometry system \citep{Acton1996} as observed by JWST accounting for light time aberration. We then sampled the ring coordinates at $0.1^\circ$ steps within the Uranian equatorial reference frame and converted those physical sky coordinates to pixel coordinates using each dither's individual WCS solution.

\begin{figure*}
    \centering
    \includegraphics[width=\linewidth]{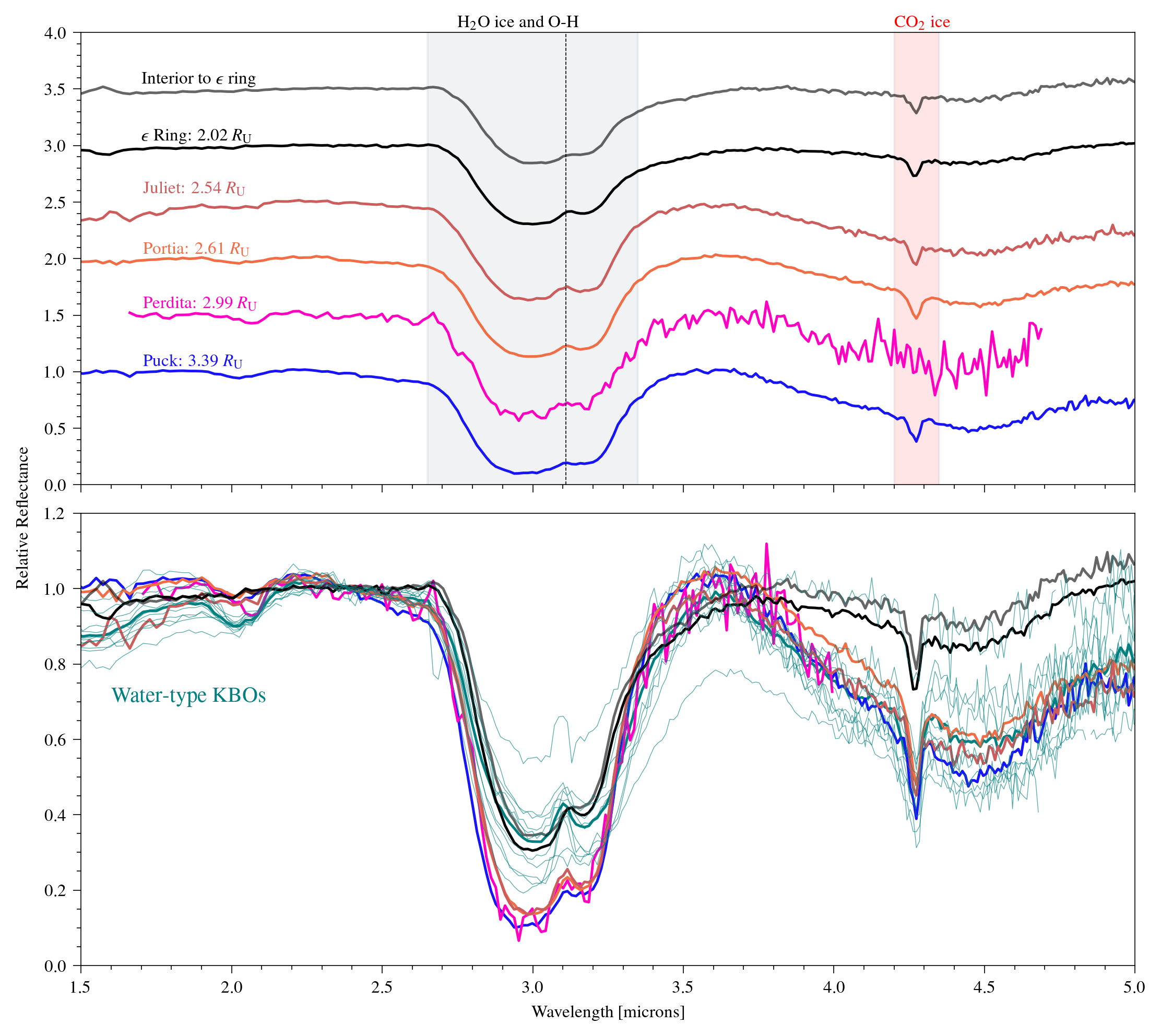}
    \caption{\textit{Top Panel:} JWST NIRSpec 1--5 \micron\ relative reflectance spectra of Uranian inner moons Puck, Perdita, Portia, and Juliet, in order of furthest to closest from the planet. Above them are the spectra of the $\epsilon$ ring and the rings interior to the $\epsilon$ ring. Each successive spectrum is offset by +0.5. There is broad similarity between all the spectra (features identified in the text), however the rings show weaker signatures of water ice and CO$_2$ than the moons. \\\textit{Lower Panel:} Comparison of the above spectra (colors correspond to the same objects as in the above panel) to water-type KBOs (in teal). The moons have simultaneously deeper 3.0 \micron\ features and shallower 2.0 \micron\ bands than the average water-type KBO, though the 4.5 \micron\ absorption is similar. The strong similarity between the two sets of spectra suggests that the small moons could have formed out of KBO-like material; possible pathways for emplacing KBO-like material at Uranus are presented in \autoref{sec:discussion}.}
    \label{fig:spec1}
\end{figure*}

For Puck, the only contribution to the background is Uranus scattered light, as there are no other moons or rings in the field-of-view. We modeled the scattered light in each wavelength slice of the 3D cube by first masking Puck using a 4-pixel radius circular aperture, and then using Astropy's \texttt{interpolate\_replace\_nans} function and a two-dimensional Gaussian kernel of 0.75 pixel width to construct the background at Puck's location on the IFU. The overall shape of the spectrum is insensitive to choice of aperture or kernel radius; we found our choices to be optimal in minimizing the effect of outlier pixels. We then subtracted out the model background, replacing the Stage2 data cube with its background-subtracted counterpart and finally extracted the spectrum using the \texttt{jwstspec} pipeline's PSF-fitting algorithm. For the other three moons, we also masked the rings when constructing the background and used a smaller 3-pixel radius circular aperture mask.

In obtaining the spectra of the rings, we sought to separate the $\epsilon$ ring from the inner rings (the faint $\lambda$ ring is included in the adjacent $\epsilon$ ring). The extended nature of these sources precluded the use of PSF-template extraction. We attempted to construct empirical PSFs of the ring, but the spatial structure of the Uranian scattered light and the location of the rings on the edge of the IFU gave unsatisfactory results. Similar model-based approaches to extract the spectra of the rings also failed. Instead, we began by masking the moons, removing pixels with high uncertainty values, or any negative science values. Taking the previously obtained coordinates of the $\epsilon$ ring, we applied an extraction aperture from the union of circular apertures of radius 1.5 pixels centered on the ring. At each wavelength slice, independently, we estimated the Uranian background from pixels outside the masked ring, which we defined to be within a 9-pixel wide annulus from the midpoint of the ring system (between $\epsilon$ and 6). The background model is given by 
\begin{equation}
    B_\lambda(r) = a_{1,\lambda}+a_{2,\lambda}\frac{50000\text{ km}}{r}+a_{3,\lambda}\left(\frac{50000\text{ km}}{r}\right)^2.
\end{equation}
We fit the three coefficients for each dither using a least-squares method after removing the 1\% brightest and faintest pixels. We then unmasked the ring and fit this model to the ring pixels, seeking to best remove the methane peaks between 1 and 1.5 \micron{} to obtain a flat near-infrared continuum. Having subtracted out the background, at each wavelength slice we took the unweighted mean of pixels in the previously constructed aperture and divided it by the solar-analog SNAP-2 (Program \#4498, Observation 19) extracted from a 2-pixel width circular aperture. The dithers were then median-combined to obtain the final spectrum. For the inner rings, we repeated this procedure, but using apertures constructed from the rings interior to the $\delta$ ring.

\section{Spectroscopic Results}
\label{sec:spec}

We show the extracted spectra of four moons, the $\epsilon$ ring, and the inner set of rings in the top panel of \autoref{fig:spec1} in order of radial distance from the planet (rings at the top, Puck at the bottom). All six spectra share most of their key spectral signatures. Most notable is the 3.0 \micron{} band associated with water ice and the O--H stretch, which appears to be deeper on the moons than the rings. Inset within the 3.0 \micron{} feature is the Fresnel reflection peak of water ice, marked by a dashed line in the figure. The Fresnel peak is indicative of the presence of grains of pure H$_2$O ice, rather than only adsorbed water \citep{Nakazawa2026arXiv}. We measured the band center of the Fresnel peak in each spectrum by fitting a second-degree polynomial to the points in the spectrum between 3.08 and 3.14 \micron{}, reporting the peak of the polynomial as the band center. Errors were computed by adding in quadrature the residuals from the fit and the uncertainty from the coarse wavelength resolution, $\Delta\lambda/\sqrt{12}$, based on the standard deviation for a continuous uniform distribution. The peak location of the Fresnel reflection is at longer wavelengths in the rings (3.116 $\pm$ 0.004 \micron{} for the $\epsilon$ ring, 3.118 $\pm$ 0.005 \micron{} for the inner rings) than on the moons (Puck: 3.105 $\pm$ 0.004 \micron{}, Portia: 3.110 $\pm$ 0.004 \micron{}, Juliet: 3.103 $\pm$ 0.004 \micron{}). The band center for the Fresnel peak traces the temperature of water ice \citep{mastrapawater, stephan_water_min11121328}. \citet{pnasbrown} measured a $\sim$3.109 \micron{} band center for the regular satellites of Uranus. Thus, our measurements are consistent with the lower temperature of the rings (77 $\pm$ 1 K, see \citealt{Molter2019AJ}) relative to the temperature of the sub-solar point of a blackbody at the distance of Uranus, which is $\sim$90 K. The 2.0 \micron{} water ice $\nu_1+\nu_3$ combination band is weakly seen on Puck, Juliet, and Portia (Perdita's noise level precludes measurement) but not on the rings, confirming previous ground-based spectroscopic measurements \citep{deKleer2013Icar}, while the $\nu_2+\nu_r$ combination lattice mode of water ice at 4.5 \micron{} is seen in all the spectra, but is much shallower in the rings.

\begin{figure*}
    \centering
    \includegraphics[width=\linewidth]{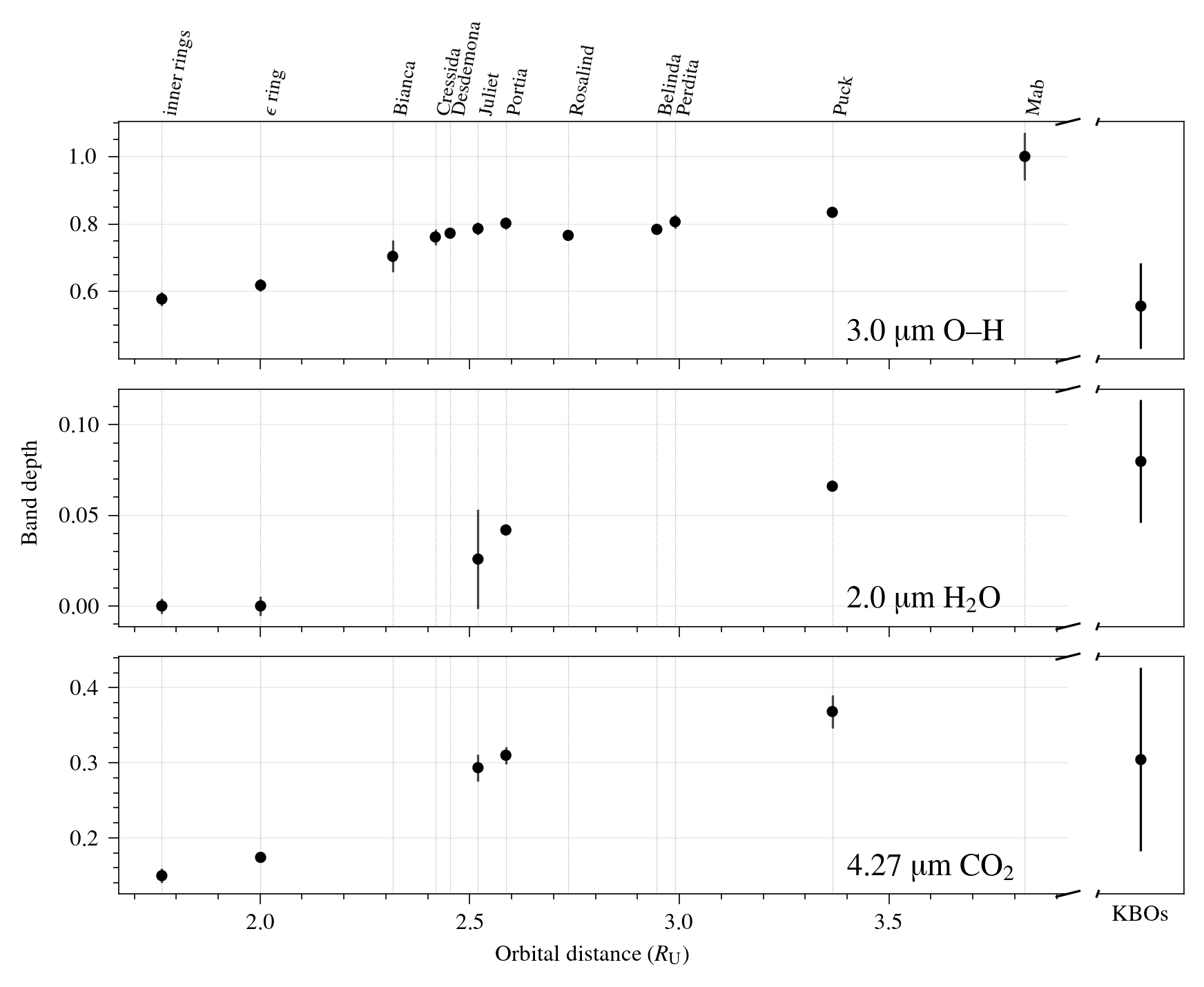}
    \caption{Band depth as a function of radial distance ($1\,R_\mathrm{U} = 25\,362\,\mathrm{km}$) for the 2.0 \micron{} water band, the 4.27 \micron{} CO$_2$ feature, and the 3.0 \micron{} absorption, in order from top to bottom. On the right side of each panel we show the corresponding band depth for water-rich Kuiper belt objects, with error bars indicating the 16 and 84-percentile measurements of the sample. All of the features show a clear trend of decreasing band depth closer to Uranus, indicative of a radially-correlated process setting the relative depths of these features. Notably, the 3.0 \micron{} band depths of the moons are significantly higher than those of KBOs, while the CO$_2$ band depth is consistent.}
    \label{fig:bands}
\end{figure*}

The moons and rings all have almost neutral near-infrared spectral slopes and do not show evidence for significant absorption by aliphatic organics at 3.4 \micron{}, which were previously suggested to be the primary darkening agent setting the low measured albedos of the moons and ring particles \citep{Soifer1981Icar,Baines1998Icar,Karkoschka2001Icar}. The lack of a 3.4 \micron{} absorption does not rule out the presence of featureless amorphous carbon, but it constitutes a strike against past suggestions of tholin-like materials comprising the rings. Instead, given the spectroscopic evidence and the measured densities of the innermost Uranian moons Ophelia and Cordelia \citep[see fig. 38 of][]{French2024Icar}, it would appear that OH-bearing minerals contribute to setting the low albedos of the rings and moons. 

CO$_2$ is present on all rings and moons, though again Perdita's SNR is too low for definitive detection. The minimum point of the band is at 4.270 \micron{}, with Gaussian fits to the continuum-subtracted feature confirming that the CO$_2$ band centers of Puck, Portia, Juliet, and the rings are consistent with 4.270 $\pm$ 0.003 \micron{}. The low spectral resolution of PRISM mode precludes obtaining a useful estimate of the band center to the third decimal point, necessary for interpretations of the exact state of the CO$_2$ \citep[see, e.g., the discussion in][]{Brown2025PSJ}. Given our measured band center, the appearance of the CO$_2$ feature is most consistent with CO$_2$ trapped in water ice \citep{Bernstein2005Icar,Galvez2008Icar} or CO$_2$ formed from irradiation of water-coated organics \citep{Gomis2005Icar}; pure crystalline CO$_2$ has a band minimum at 4.267 \micron{}, which may be somewhat too short \citep{Hudson_2025}. The observed band center is longer than what arises from CO$_2$ trapping in minerals such as carbonates or hydrated silicates \citep{Hibbitts2007Icar,Pandya2026PSJ}. Bodies such as Phoebe and many of the smaller water-rich Kuiper belt objects have similar CO$_2$ band centers to the Uranian rings and small moons \citep{PinillaAlonso2025NatAs,Belyakov2025PSJ, Wong2025PSJ}, while numerous other bodies such as the Jupiter Trojan Eurybates or larger objects in the Kuiper belt like Charon have shorter-wavelength CO$_2$ band centers \citep{Wong2024PSJ, Protopapa2024NatCo, Wong2025PSJ}. The available data do not provide a conclusive verdict on the state of CO$_2$ in the rings or small moons. However, we note that the appearance of the band is highly distinct from the CO$_2$ on the large regular satellites, where it has a much more prominent spectral appearance \citep{2026arXiv260705600C}. We note that a 3.9 \micron{} feature previously reported in presentations while this work was in progress and as referenced in \cite{2026arXiv260705600C} does not appear with optimal background subtraction, and was likely an artifact. 

\subsection{Comparisons to Kuiper Belt Objects}
In the lower panel of \autoref{fig:spec1}, the spectra of the moons and rings are overlaid on top of water-type KBOs, illustrating their overall similarity. The obvious spectral match between the small moons and rings and KBOs invites speculation on the formation history and origin of the Uranian system; however, we leave this discussion for \autoref{sec:discussion}. The subtle differences between the spectra in the Uranian system and Kuiper belt objects could be a result of differences in grain size. The shape of the 3.0 \micron{} feature in both sets of objects is very similar, yet the strength of the 2.0 and 4.5 \micron{} features relative to the 3.0 \micron{} is lower on the moons and rings than on the KBOs despite their similar albedos of $\sim$0.05 to 0.1, consistent with the moons and rings of Uranus hosting finer-grained water ice than the KBO surfaces \citep[see, e.g.][]{Wiscombe1980JAtS, Clark1981JGR, Hansen2004JGRE}. One possibility is that the delivery of sub-micron sized grains from the outer moons (in analogy to Saturn's E ring) could be responsible for the difference in the appearance of the water ice bands. Additionally, the highly collisional environment of the ring-moon system may have altered the grain size distribution over long timescales. However, it is worth noting that micron-sized grains are not a major constituent of the dense rings, including the $\epsilon$ ring \citep{Ockert1987JGR}. Therefore it is unclear if collisional grinding has contributed to altering the grain size distribution at spectrally relevant sizes. 

We also quantify the similarity between the Uranian ring-moon system and the water-type KBOs. Using the Mahalanobis distance, which is the multivariate version of the commonly used $z-$score for the normal distribution, we can estimate the probability that our objects belong to the space of water-type KBO spectra. We use a three-dimensional space of 2.0 \micron{} band depth, Fresnel peak height, and CO$_2$ band depth to estimate this distance. We leave out the 3.0 \micron{} feature depth, because the ratio of  the 2.0 \micron{} to 3.0 \micron{} bands is potentially set by grain size, rather than abundance. For Puck, Portia, and Juliet, we find that they are statistically indistinguishable from random draws from the KBO population ($p=0.9, 0.4, 0.2$ respectively). The rings, however, due to their lack of a 2.0 \micron{} feature and weaker Fresnel peaks have $p<0.05$, and thus are not statistically similar to the KBOs. This result is expected, as the physics of ring-moon cycling, differences in particle size, and effects of the Roche limit (discussed below), will invariably lead to spectral differences from the basic water-type KBO spectral shape. Finally, we take the set of water-type, organics-type, and CO$_2$ type KBOs (see \citealt{Belyakov2026SciA}), and use them to try and fit the spectrum of Puck with a non-negative least squares algorithm. We then compare the quality of each model fit with the Akaike information criterion (AIC = $2k- 2\log{\hat{L}}$ where $k$ is the number of parameters, $\hat{L}$ the maximum likelihood \citealt{Akaike1974ITAC}), with the $\Delta$AIC providing the criterion for model selection. Fitting Puck's spectrum with water-rich KBOs is highly favored, with the $\Delta$ AIC $>10^4$ for both Organics-type and CO$_2$-type vs. water-type fits.

\subsection{System Trends}
We measure the band depths of 3.0 \micron{}, 2.0 \micron{} and CO$_2$ in order to show radial trends in the system. For the 2.0 \micron{} water band and the 4.27 \micron{} CO$_2$ feature we continuum-divide and fit a Gaussian, with residuals on the fit as the source of the error bars. The rings show no significant 2.0 \micron{} band. For the 3.0 \micron{} band depths, we combined our measurements with those available from JWST spectrophotometry measured by \citep{Hedman2025PSJ} -- where available, we report the band depth measured by spectroscopy rather than spectrophotometry. For our spectra, we convolve the JWST NIRCam photometric bands F210M and F300M with the spectrum, and then report the band depth as one minus the ratio of the reflectance in the two filters. Mab, which has no reported 3.0 \micron{} detection in \citep{Hedman2025PSJ}, is assumed to have zero reflectivity at 3.0 \micron{}, which is a reasonable assumption given its presumed similarity to Miranda and the larger Uranian satellites based on its very blue near-infrared slope. We repeat all the band depth measurements for the water-type KBO sample, with errorbars showing the 1$\sigma$ ranges across the sample. We show the band depth trends in \autoref{fig:bands}. Excluding Mab from the 3.0 \micron{} trend, the correlation coefficients for the 2.0 \micron{}, 3.0 \micron{}, and 4.27 \micron{} features with radial distance are $r=0.97, 0.90, 0.96$ respectively, with $p<0.01$ for all three samples. The presence of these trends in the inner moon system may be linked to the concept of ring-moon cycling, whereby collisions between moons form rings, out of which small satellites can then accrete \citep{Hesselbrock2017NatGe, Hesselbrock2019AJ, Cuk2022AJ, Ward2025MNRAS}. After moonlets have re-accreted, the material remaining in the rings represents particles and clumps of density similar to or under the local Roche density limit. Thus, material is segregated over time by density within the system, which could explain the observed compositional trends.   

\section{Origin of the Rings and Ringmoons}
\label{sec:discussion}

Spectral coverage by JWST of the water ice and CO$_2$ absorption features in the infrared has enabled significant progress towards understanding the compositions of the many satellites of the ice giant planets. This Letter follows the publication of several key results on the Uranian and Neptunian systems, which we briefly summarize to facilitate the interpretation of our data. 

First, the Neptunian inner moons, characterized in \cite{davismoons}, are strikingly different from water-rich Kuiper belt objects and are likewise unlike the Uranian inner moons. Neptune's inner moons Larissa and Galatea, the latter of which is of similar size to Puck, show no water ice or CO$_2$, but instead display phyllosilicates on their surfaces. Hydrated minerals require prolonged exposure to higher temperature and pressure conditions, presumably in the mantles of larger, differentiated satellites. Neptune's present-day satellite system is thus a product of Triton destroying a once-larger satellite system, whose re-accreted remnants are the present-day inner moons of Neptune \citep{Belyakov2026SciA, davismoons}. Puck's spectrum is decidedly not like that of Neptune's moon Galatea, perhaps suggesting distinct origins and explanations for the small ring and moon systems of the two planets. Secondly, recent measurements of the D/H ratio on the regular moons of Uranus by \cite{pnasbrown} have revealed a cometary or Kuiper belt-like D/H ratio. The signature is consistent on all the moons except Miranda, where it may be slightly elevated in comparison to Ariel, Umbriel, Titania, and Oberon. Thus, any explanation of the formation history of the Uranian satellite system must be consistent with these D/H measurements, which have already ruled out satellite formation through an impact-generated vapor disk \citep[see the discussion in][]{pnasbrown}. Finally, in this Letter, we report that the rings and inner moons of Uranus are of Kuiper belt-like material, specifically resembling the water-type KBOs, and that there exists a radial trend in the abundance of volatiles within the ring-moon system. We now discuss the implications of these results for possible formation scenarios of the Uranian satellite system, and specifically that of the small moons and rings.

\subsection{Destruction and Re-accretion of a First-Generation Satellite System}
One of the primary hypotheses for explaining the equatorial orbits of the Uranian satellites is that the collision which tilted Uranus disrupted the planet's original satellites, which then re-accreted in the new, tilted equatorial plane \citep{Morbidelli2012Icar}. These primordial Uranian satellites must have formed out of solids accreted from the protosolar nebula to be consistent with the elevated D/H signature. We find that it is difficult to explain the inner moons of Uranus forming out of material from a first generation of satellites or from a \cite{Canup2006Natur} style gas-starved disk. Forming, tilting, and then re-accreting the small satellites out of circumplanetary material is hard to reconcile with their KBO-like spectra; material in the circumstellar and circumplanetary environments is not the same, with the planet inducing temperatures in the disk above several hundred Kelvin at distances of a few planetary radii \citep{Canup2006Natur, Batygin2020ApJ}. 

Had the original satellites begun differentiation prior to the tilting event, we would potentially observe the same phyllosilicate-like signatures on the inner moons as seen on Larissa. In this scenario, the Uranus-tilting impact and disruption of satellites should happen early, to prevent mantle-like material from large satellites accreting into the inner moons and rings, as seen at Neptune \citep{davismoons}. Another factor to consider is that Miranda's distinct properties (low density, higher D/H) as compared to the four other regular satellites have previously been suggested to be the result of a distinct formation from impactor material in a disk formed during the Uranus-tilting collision \citep{Hesselbrock2019AJ, Salmon2022ApJ}. If the impactor was a large water-type KBO, similar to, for example, 2002 MS4, it is possible that the result would be a disk that could produce Miranda and a compact inner satellite system. Simulations of such a disk must then be able to form the inner moons out of undifferentiated, KBO-like material. A measurement of the deuterium to hydrogen ratio in the rings would help further constrain formation scenarios and determine whether the inner part of the system formed concurrently with Miranda.

\subsection{Formation from tidally disrupted Outer Solar System material}
A distinct method of forming satellites around the giant planets is by forming a ring of tidally disrupted material, described first for Saturn's rings in \cite{Charnoz2010Natur} and subsequently for all the giant planets in \cite{Crida2012Sci}. In this scenario, comets and Kuiper belt objects that happen to pass within the ice Roche limit of a giant planet are torn apart, forming a ring of debris, which then spreads through viscous and tidal evolution, allowing for satellite formation in configurations similar to those observed at present. In this scenario, the D/H ratio constraint is satisfied, as discussed in \cite{pnasbrown}. The compositions of the inner moons are a natural consequence of this formation story, and, as noted in \cite{Crida2012Sci}, the present-day orbital configuration of the Uranian moons with respect to their assumed individual masses follows the predictions from accretion out of an early massive ring of disrupted material. 

A final explanation for the observed data is that the regular satellites and the ring-moon system are simply unrelated, and that the small inner moons and the rings arise from the later tidal disruption of a water-type Kuiper belt object. The processes previously invoked to explain the dynamics and D/H ratio of the regular satellites can be divorced from discussion of the inner moons and rings. Even if only a few percent of a disrupted body's material is incorporated into the ringmoons, a $\sim$500 km KBO would have supplied enough material to form the present-day Uranian small moons and rings.

The chaotic dynamics of the inner satellite system have been used to argue that the Uranian rings and small moons are not ancient (see the discussion in the supplementary materials of \citealt{Crida2012Sci}). Notably, studies of the chaotic evolution of the inner system have, over the years, pushed back estimates of the collisional timescales in the system to at least 10$^8$ years \citep{French2015AJ, Cuk2022AJ}. Our detection of radial sorting of volatile-poor to volatile-rich material as a function of distance from Uranus may require a long cycling of material between rings and moons. Thus, the inner component of the Uranian system need not be young to have preserved material in small moons to the present day. If the inner moons do indeed postdate the regular satellites, they may only be marginally younger, and would have been able to form from scattering planetesimals available after the early Solar System's dynamical instability.

\section{Summary}
We recapitulate the key findings of this Letter, which presented 1-5 \micron{} spectroscopy of the inner moons and rings of Uranus and connected their surface composition and possible origins to the water-type Kuiper belt objects.

\begin{itemize}
    \item Our methods highlight several JWST pipeline settings that improve the SNR of NIRSpec spectra for faint point-source targets. We recommend using the EMSM weighting method for cube building, applying the non-linear wavelength solution, and relying on a median fit rather than an FFT fit for $1/f$ readnoise noise correction.
    \item The Uranian rings and moons all show evidence for water ice (via the Fresnel peak) and CO$_2$. Nonetheless, the rings do not have the 2.0 \micron{} or much of a 4.5 \micron{} feature of water ice, perhaps suggesting lower abundance and/or smaller grain size than on the small moons, which show both of these features.
    \item There is no apparent sign of absorption features in the 3.2--3.5 \micron{} range from aliphatic or aromatic organics. Given the deep 3.0 \micron{} feature in the rings, as well as the high density ($\sim1.5$ g/cc) of Cordelia, which orbits adjacent to the $\epsilon$ ring \citep{French2024Icar}, it is likely that the material responsible for the ring-moon system's low albedo is, in part, some OH-bearing silicate.
    \item Water-rich Kuiper belt objects, hypothesized to be the planetesimals that formed closer to the giant planet region than the Cold Classical KBOs \citep{PinillaAlonso2025NatAs}, are a very close spectral match to the ring-moon system. We use this similarity to suggest that the inner Uranian moons are remnants of a tidally-disrupted KBO. 
    \item Reconciling the spectral similarity between Puck and the water-type KBOs is more difficult in a scenario where the inner moons are the destroyed remnants of first-generation satellites. Such a scenario is strongly suggested to be the case at Neptune \citep{Belyakov2026SciA, davismoons}. However, the Uranian moons appear spectrally distinct from the Neptunian ring-moon system's phyllosilicate-rich spectra.
    \item The hypothesized ring-moon cycling \citep{Hesselbrock2019AJ}, wherein collisions between moons form rings, which then re-coalesce into moons, may be evidenced by the radial sorting of material in the system. The features of CO$_2$ and H$_2$O both show decreasing band depth closer to Uranus.
\end{itemize}

\begin{acknowledgments}
This work is based on observations made with the NASA/ESA/CSA James Webb Space Telescope. The data were obtained from the Mikulski Archive for Space Telescopes at the Space Telescope Science Institute, which is operated by the Association of Universities for Research in Astronomy, Inc., under NASA contract NAS 5-03127 for JWST. The JWST/NIRSpec observations are associated with Program \#4645. The specific observations analyzed can be accessed via \dataset[doi: 10.17909/0a2k-tq09]{https://doi.org/10.17909/0a2k-tq09}. We thank Imke de Pater, Matt Hedman, Tracy Becker, Shawn Brooks, Richard Cartwright, Tilmann Denk, Richard Jerousek, and Maryame El Moutamid for valuable insights on the Uranian ring-moon systems. We also would like to acknowledge all the attendees of the 2025 JWST Solar System Workshop in Meudon, France for discussions that contributed to this work and our other publications involving data obtained as part of JWST program \#4645.
\end{acknowledgments}

\bibliography{bibliography}{}
\bibliographystyle{aasjournal}

\end{document}